\documentstyle[prb,aps,epsf,twocolumn]{revtex}
\begin{document}
\draft

\title{The structural and electronic properties of germanium clathrates}

\author{Jijun Zhao, Alper Buldum, Jianping Lu}

\address{Department of Physics and Astronomy, University of North Carolina
at Chapel Hill
}

\author{C.Y.Fong}
\address{Department of Physics, University of California at Davis\\
}

\date{\today}
\maketitle
\begin{abstract}

The structural and electronic properties of germanium clathrates Ge$_{46}$
and K$_8$Ge$_{46}$ are studied by first principles calculations within the
local density approximation. The equilibrium structures are obtained by
{\em ab initio} pseudopotential calculation combined with dynamic
minimizations. The clathrate structure is found as a low energy phase
for germanium. The electronic band structures for Ge$_{46}$ clathrates
are calculated and the band gap is found to be considerably larger than that
of the diamond phase. Due to the effect of pentagonal rings, strong similarity
in electronic properties between clathrate and Ge$_{24}$ fullerene structure
are found. The effect of doping clathrate cages with metal atoms are examined.
The K$_8$Ge$_{46}$ clathrate is found to be metallic with the
conduction bands only slightly modified by K dopants.

\end{abstract}


\narrowtext

\section{Introduction}

The Si and Ge clathrates can be viewed as covalent fullerene solids which is
composed of three-dimensional networks of fullerene cages connected by face
sharing. The silicon clathrate compounds M$_x$Si$_{46}$ and M$_x$Si$_{136}$
(M$=$Na, K, Rb, and Cs) were first synthesized in 1965 \cite{1}. The
structures of semiconductor clathrates can be classified into two cubic
structural types with 46 atoms or 136 atoms per unit cell \cite{1}. As shown
in Fig.1, the type I clathrate (Clathrate-46) is formed by two smaller
14-face pentagonal dodecahedra (12 five-fold rings, $I_h$) and six larger
16-face tetrakaidecahedra (12 five-fold rings and 2 six-fold rings, D$_{6d}$).
The structure of type II clathrate (Clathrate-136) consists of sixteen
smaller pentagonal dodecahedra (12 five-fold rings, $I_h$) and eight larger
hexakaidecahedra (12 five-fold rings and 4 six-fold rings, $T_d$). Electronic
structure calculations based on different methods have shown that these
two open network structures have similar electronic properties \cite{2,3}.

\begin{figure}
\centerline{
\epsfxsize=3.0in \epsfbox{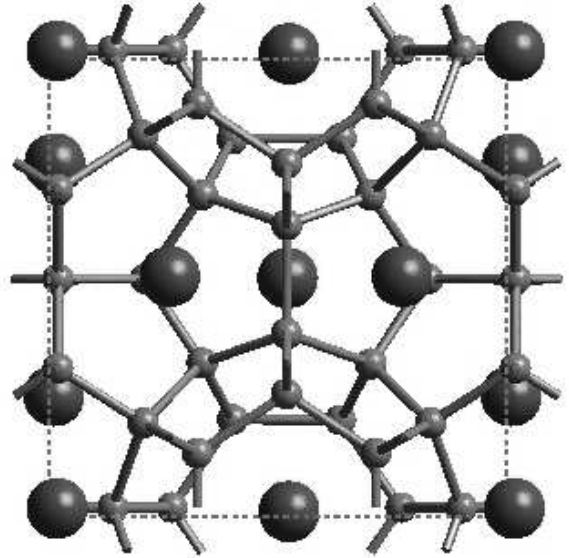}
}
\caption{Atomic structure of fully-relaxed K$_8$Ge$_{46}$ clathrate.
Lattice constant of the simple cubic cell is 10.45 $\AA$.
}
\end{figure}

The research interests of semiconductor clathrates come from several aspects:
(1) possible alteration of electronic structures and energy gap from standard
diamond form \cite{2,3}, (2) metal-insulator transition in M$_x$Si$_{136}$
with different concentration of metallic impurity \cite{4,5}, (3) the
finding of superconductivity behavior in Na$_x$Ba$_y$Si$_{46}$ \cite{6},
(4) candidate for thermoelectric applications \cite{7}, (5) template to
from three dimensional arrays of nanosized clusters \cite{8}, (6) similarity in
structural and electronic properties between semiconductor clathrates and
nanoclusters \cite{9,10}.

There are lots of the experimental and theoretical studies on
Si$_{46}$ and Si$_{136}$ clathrates and their compounds
\cite{1,2,3,4,5,6,7,8,9,10,11,12,13,14,15,16,17,18,19,20,21}. The structural,
electronic and vibrational properties of Si clathrates are investigated at
various theoretical levels ranging from {\em ab initio} \cite{2,9,11,12} to
tight-binding \cite{3,13} and empirical potential \cite{14}. The most
interesting result from those calculations is that the band gap is about
0.7 eV higher than that of diamond phase. Experimental works on Si clathrates
include resistivity and magnetization \cite{6}, transport properties \cite{7},
photoemission spectroscopy \cite{10}, NMR \cite{15,16,17}, Raman \cite{18,19},
ESR \cite{20}, neutron scattering \cite{21} etc.

In contrast to the intensive studies on silicon clathrates, our current
knowledge on germanium clathrates is rather limited. Recently, the germanium
clathrate compounds such as K$_8$Ge$_{46}$, Rb$_x$Ge$_{46}$,
Na$_x$Ge$_{136}$, Cs$_8$Na$_{16}$ have been synthesized and their structures
are analyzed with X-ray diffraction \cite{22,23}. An empirical potential
calculation has been performed on pure germanium clathrates
\cite{14}. However, there is no first principles electronic structure
calculation and their fundamental electronic properties are still unclear
theoretically. In this work, we report results of first principles study on
the structures and electronic properties of Ge$_{46}$ and K$_8$Ge$_{46}$
clathrates. The equilibrium structure, electronic band and band gap,
electronic density of states and electron density distribution are obtained
and discussed.

\section{Computational methods}

First principles SCF pseudopotential method \cite{24,25,26} is used to
perform static calculation on the electronic structures and total energy of
Ge$_{46}$ and K$_8$Ge$_{46}$ in ideal clathrate structures with different
lattice constants. The ion-electron interaction is modeled by numerical BHS
norm-conserving nonlocal pseudopotential \cite{27} in the Kleinman-Bylander
form \cite{28}. The Ceperley-Alder's exchange-correlation parameterized by 
Perdew and Zunger is used for the LDA in our program \cite{29}. 
The kinetic energy cutoff for plane-wave basis is chosen as 12
Ryd. Ten symmetric {\bf k} points generated in Monkhorst-Pack spirit
\cite{30} are employed to sample the Brillouin zone.

From the static calculation, the equation of states with ideal crystal
structures and the equilibrium lattice constants are obtained. These
ideal structures are further optimized by using structural
minimization via conjugate gradient technique (CASTEP \cite{31}).
The CASTEP program is based on plane-wave pseudopotential technique.
It can relax the atomic position by computing the force
acting on atoms from electronic calculation and moving atoms efficiently
in a numerical way \cite{25}.

\section{The structures and band gaps of {\mbox{Ge$_{46}$}}}

In Fig.2, we present the equation of states for both Ge$_{46}$ clathrate and
diamond phase that is obtained from first principles static calculations.
We find that the Ge$_{46}$ clathrate is a locally stable structure and its
energy is only about 0.08 eV per atom higher than that of diamond phase. For
comparison, the $\beta$-tin phase of germanium is about 0.25 eV higher in
energy than the diamond phase \cite{24}. The low energy feature of clathrate
phase is a natural consequence of its four coordinate characteristics and may
be partially attributed to the softness of the bond-bending distortion modes
\cite{9}. As compared to diamond structure, the volume per atom in the
clathrate phase is increased by about 14.8$\%$. All of these results are very
close to the previous empirical potential simulation on Ge$_{46}$, in which
the change of atomic volume in type I clathrate is 15.3 $\%$ and its energy is
0.071 eV per atom higher than diamond \cite{14}.

\begin{table}
Table I The structural properties and band gaps of perfect and relaxed
clathrate compared with previous LDA local orbital (LDA-LO) and
tight-binding molecular dynamics (TBMD) calculation on Si$_{46}$ clathrate.
V/V$_0$ is the reduced atomic volume. $l$ denotes the Ge$-$Ge bond length.
$a$ is the equilibrium lattice constant. $E_g$ is the electronic band gap
(indirect gap for both diamond and clathrates). The numbers in brackets are
experimental values \cite{33,22}.
\begin{center}
\begin{tabular}{l|l|l|l|l}
                          & V/V$_0$ & $l$ (\AA) & $a$ (\AA)      & $E_g$ (eV) \\ \hline
Ge Diamond                & 1.00    & 2.407             & 5.56 (5.66)  & 0.40 \\
Ge$_{46}$ (ideal)         & 1.148   & 2.375, 2.540  & 10.43        &  1.25         \\
Ge$_{46}$ (relaxed)       & 1.128   & 2.380 $\sim$ 2.433     & 10.37        &  1.46    \\
K$_8$Ge$_{46}$ (ideal)    & 1.271   & 2.562,  2.739     & 10.79         &   --        \\
K$_8$Ge$_{46}$ (relaxed)  & 1.155   & 2.346 $\sim$ 2.492     & 10.45 (10.66) &   --   \\ \hline
Si$_{46}$ (LDA-LO)\cite{2}& 1.17    &  2.38 $\sim$ 2.43      & 10.35         &  2.50 \\
Si$_{46}$ (TBMD)\cite{3}  & 1.14    &  2.367            & 10.20         &  1.82      \\
\end{tabular}
\end{center}
\end{table}

In previous theoretical study on bulk silicon and germanium of various phases
\cite{24}, considerable similarity in the bonding behavior and phase diagram
are found between silicon and germanium. In Table I, we summarize our results
on the structural properties and band gaps for perfect and relaxed Ge$_{46}$
and K$_8$Ge$_{46}$ clathrate and compare them with the previous LDA and
tight-binding calculations on Si$_{46}$ clathrate \cite{2,3,9}. We find the
difference of volume and energy between clathrate and diamond phase for
germanium and silicon are comparable.

\begin{figure}
\centerline{
\epsfxsize=3.0in \epsfbox{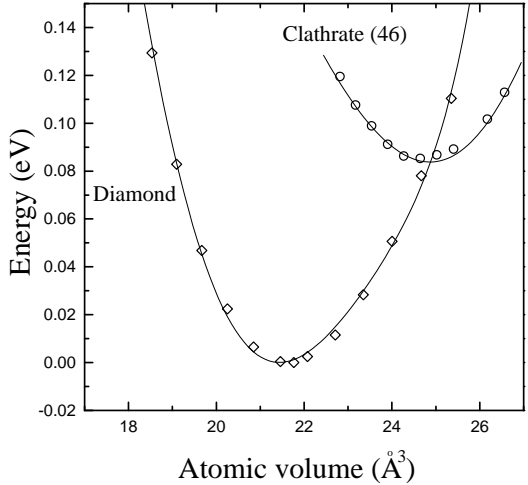}
}
\caption{Energy per atom of diamond and Ge$_{46}$ clathrate as a function of
atomic volume.
}
\end{figure}

The ideal clathrate structure for the Ge$_{46}$ at equilibrium lattice
constant is further optimized with CASTEP plane-wave pseudopotential
calculation. Both atomic positions and unit cell parameters are allowed to
relax. The SCF pseudopotential code used in static calculation has also
been used to test the total energy for the initial and final structure
and the results agree with CASTEP calculations. After optimization,
the lattice constant of simple cubic unit cell decreases from 10.43 $\AA$ to
10.37 $\AA$. The relative atomic positions are only slightly relaxed from their
initial configuration. The range of bond length distribution has also been
narrowed, i.e., from the 2.375 $\AA$ $\sim$ 2.540 $\AA$ in ideal clathrate
structure to 2.38 $\AA$ $\sim$ 2.433 $\AA$.

\begin{figure}
\centerline{
\epsfxsize=3.0in \epsfbox{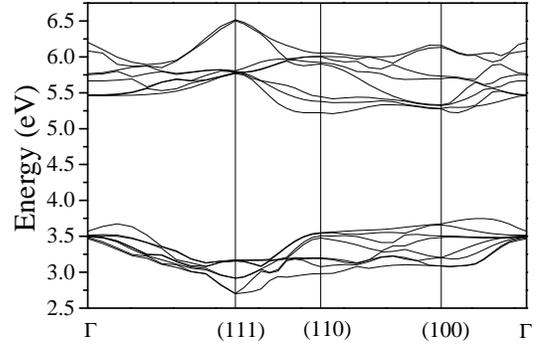}
}
(a)
\centerline{
\epsfxsize=3.0in \epsfbox{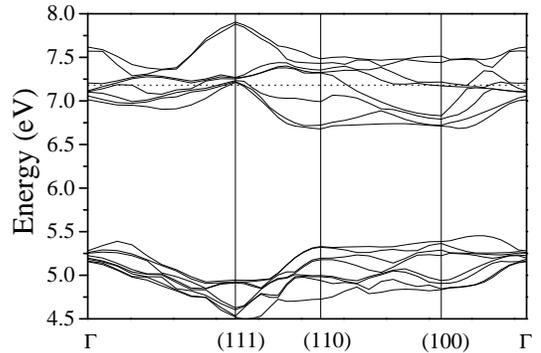}
}
(b)
\caption{Electronic band structure for (a) Ge$_{46}$, (b) K$_8$Ge$_{46}$
clathrate near Fermi level.}
\end{figure}

The electronic band structure calculations are carried out for Ge$_{46}$
clathrate with the equilibrium structure. The near band-gap band structures
for the minimum energy configurations of Ge$_{46}$ clathrates are plotted in
Fig.3(a). Both the valence band maximum and conduction band minimum are
located on the $\Gamma$ to $X$ line and they are very close in {\bf k} space.
An indirect gap of 1.46 eV is found. As for bulk germanium in diamond phase,
our calculations have yielded an indirect band gap of
0.40 eV between $L$ and $\Gamma$ points. The underestimation of band gap is a
common feature of LDA calculations. Nevertheless, our current results suggests
that the band gap of Ge$_{46}$ phase is about 1 eV higher than that of the
diamond phase, which is similar to the 0.7 eV increment in band gap
found for Si$_{46}$ and Si$_{34}$ clathrates \cite{2,3,9,10}. Furthermore,
the electronic band structure of Ge$_{46}$ clathrate are calculated
independently via a simple tight-binding model. The tight-binding hopping
parameters for germanium are taken from Ref.[32] and the parameters
for nearest neighboring atoms in clathrate structure is assumed to be the
same as that in diamond. According to tight-binding calculation, the band
gap of diamond is 1.13 eV and the gap for relaxed clathrate Ge$_{46}$ is
2.93 eV. Although the tight-binding method usually overestimate the band
gap of a system, the increment of 1.80 eV in the band gap from tight-binding
model is in reasonable agreement with {\em ab initio} calculation. Therefore,
we expect that the true increase of band gap $\Delta$ from diamond to Ge$_{46}$
clathrate phase might be between the LDA and TB prediction, i.e.,
1.06 eV $\leq$ $\Delta$ $\leq$ 1.80 eV. This result suggests that the clathrate
materials might be useful in the new electronic and optical application in the
future.

Since the Ge$_{46}$ clathrate structure is essentially a 3D network composed
by Ge$_{20}$ and Ge$_{24}$ cages with face sharing, it is interesting to
compare the electronic properties of the individual Ge$_{20}$ and Ge$_{24}$
cages with that of the clathrate. SCF pseudopotential electronic structure
calculations on isolated Ge$_{20}$ and Ge$_{24}$ clusters in fullerene cages
are performed. The clusters are placed in a large simple cubic supercell with
length of 28 a.u. In Fig.4, we present the calculated electronic density of
states (DOS) for Ge$_{20}$ and Ge$_{24}$ cages along with DOS for diamond and
Ge$_{46}$. The detailed analysis are given in the following.

\begin{figure}
\centerline{
\epsfxsize=3.0in \epsfbox{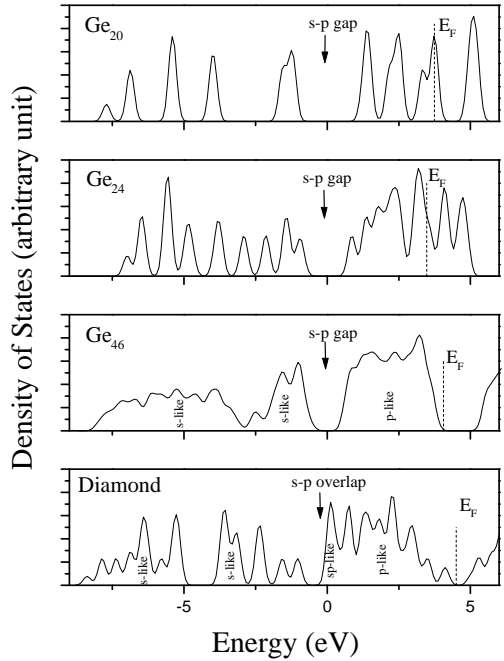}
}
\caption{Electronic density of states (DOS) of Ge$_{20}$, Ge$_{24}$
fullerene, bulk germanium in Ge$_{46}$ clathrate and diamond phases
Gaussian broadening of 0.136 eV is used.
}
\end{figure}

Firstly, we find that most of the peaks in the DOS of Ge$_{46}$ can be
assigned to Ge$_{24}$, while the DOS of Ge$_{20}$ also show some similarity
with Ge$_{46}$. The DOS of Ge$_{24}$ is closer to that of clathrate
because most atoms in clathrate are associated with Ge$_{24}$ cage.
The presence of s-p gap in due to the large number of five-member rings in the
cage and clathrate structure. This remarkable similarity implies that the
Ge$_{46}$ clathrate can be taken as the small Ge fullerene assembled solid with
both geometrical and electronic hierarchy. However, both the Ge$_{24}$ and
Ge$_{20}$ cages are not semiconductor systems and do not have open band gaps
like Ge$_{46}$ clathrates. The band splitting in Ge$_{46}$ can be attributed to
the sharing of face atoms by Ge$_{20}$ and Ge$_{24}$ and the interaction
between neighboring fullerene cages.

On the other hand, we can compare the DOS of clathrate and diamond in Fig.4.
Several significant differences can be found. Besides the improvement of band
gap in clathrate upon diamond phase, the total width of valence band of
Ge$_{46}$ (about 12 eV) is narrower than that of diamond (about 13 eV). This
phenomenon has been also predicted in Si$_{46}$ clathrate \cite{9} and observed
experimentally \cite{10}. We also found a gap opening between the s-like
and p-like states in the case of Ge$_{46}$ clathrates as well as Ge$_{20}$,
Ge$_{24}$ clusters, while $s$ and $p$ states are overlapped in the DOS of
diamond. These remarkable differences in the electronic structure of
clathrate and diamond can be understood by the large portion of five-fold
ring in clathrate structure. In the diamond lattice consisting of $100\%$
six-fold ring, the $4s$ orbital can form complete antibonding states, which can
distribute to high energy and overlap with the lower $4p$-like states.
In contrast to diamond, the Ge$_{46}$ clathrate is composed of $87\%$
five-fold rings and $13\%$ six-fold rings. In the five-fold ring, $4s$ orbital
cannot form complete antibonding states so that the top of $4s$-like states
is still lower in energy than the bottom of $4p$-like and a gap between $s$
and $p$ states inside valence band opens. The similar effect of five-fold
ring on the $4p$ bonding orbitals will induce the incompleteness of $4p$-like
states. As a consequence, the valence band top of Ge$_{46}$ is lower than that
of diamond, which corresponds to the narrowing of valence band width and
broadening of fundamental gap between valence and conduction band.
The novel electronic properties caused by five-fold ring and the similarity
between clathrate and fullerene cages may be explored for future
electronic and optical applications.

\section{Electronic structures of {\rm K$_8$Ge$_{46}$} clathrate}

Although the pure silicon and germanium semiconductor clathrates are predicted
to be locally stable, experiments have only synthesized the metal-doped
clathrate compounds of Si and Ge. The metallic impurity atoms inside the
clathrate might influence the electronic properties of clathrate due to the
interaction between the metal atoms and semiconductor skeleton.
Here we have chosen the K$_8$Ge$_{46}$ clathrate as such a model system to
study the doping effect on germanium clathrate.

\noindent
The structural properties of K$_8$Ge$_{46}$ are studied with the same procedure
we have applied on Ge$_{46}$ clathrate: a static calculation of equation of
states followed by a full relaxation. The minimized structure of
K$_8$Ge$_{46}$ clathrate is shown in Fig.1 and the structural parameters
of this minimum energy configuration is given in Table I. It is natural to
find that the lattice constant of cubic unit cell for K$_8$Ge$_{46}$ clathrate
is larger than pure Ge$_{46}$ clathrate in both ideal and relaxed structure
because of the K atoms inside the fullerene cage. We also find a
decrease of lattice constant from 10.79 $\AA$ to 10.45 $\AA$ after structure
minimization. The minimized lattice constant (10.45 $\AA$) for the
K$_8$Ge$_{46}$ is in good agreement with experimental value (10.66 $\AA$)
\cite{22}. During the minimization, the geometries of K$_8$Ge$_{46}$ relax
from the perfect clathrate network a little more than that happened in the
Ge$_{46}$.

First principles SCF pseudopotential electronic structure calculation
has been performed on the relaxed K$_8$Ge$_{46}$ clathrate. In Fig.3(b), we
presented the electronic band structure of K$_8$Ge$_{46}$ near Fermi level.
The system is found as metallic due to the K dopants. We can further examine
the highest valence bands and lowest
conduction bands of Ge$_{46}$ and K$_8$Ge$_{46}$ shown in Fig.3(a) and (b)
in detail. The valence bands of K$_8$Ge$_{46}$ are very closed to the
original bands in Ge$_{46}$, while the conduction band structures
have been slightly modified upon the inclusion of K atoms.

\begin{figure}
\centerline{
\epsfxsize=3.0in \epsfbox{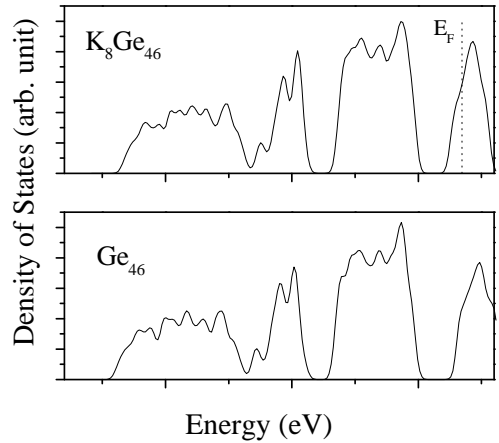}
}
\caption{Electronic density of states (DOS) of Ge$_{46}$ and K$_8$Ge$_{46}$
clathrate. Gaussian broadening of 0.136 eV is used. Note the similarity in
the two DOS.
}
\end{figure}

We also study the difference of the electronic properties between
the pure and doped systems by comparing their the electronic density of
states (DOS). The density of states for Ge$_{46}$ and K$_8$Ge$_{46}$ clathrate
are compared in Fig.5. The DOS of valence electrons in K$_8$Ge$_{46}$ is very
close to that in Ge$_{46}$ while the DOS for conduction electrons of
K$_8$Ge$_{46}$ are somewhat different from that in Ge$_{46}$. On the other
hand, the gap between valence and conduction band is 0.23 eV narrower in the
case of K$_8$Ge$_{46}$.

\begin{figure}
\centerline{
\epsfxsize=3.0in \epsfbox{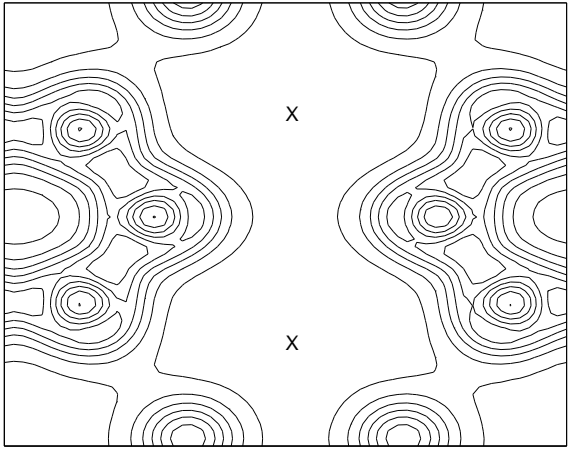}
}
\caption{Contour plots of the electron densities of K$_8$Ge$_{46}$ on the
(100) plane. No charge density is found at K site (marked by X), indicating
complete charge transfer.
}
\end{figure}

The analysis on the charge transfer and chemical bonding effects
have been presented in the contour plots of the electron densities of
K$_8$Ge$_{46}$ on the (100) plane in Fig.6. We have also
calculated the charge density distribution of Ge$_{46}$ and find it is very
close to that of K$_8$Ge$_{46}$. As shown in Fig.6, there is almost no charge
sitting on K sites.
This result is consistent with previous calculation on Na$_2$Ba$_6$Ge$_{46}$
\cite{9}, which found a rather simple charge transfer from Na to the Si skeleton.
In their calculation, some hybridization are found between Ba and Si since Ba
atoms has some low lying $5d$ orbitals \cite{9}. This difference also corresponds
to the DOS and other electronic properties. A high DOS peak at Fermi energy is
found for Na$_2$Ba$_6$Ge$_{46}$ \cite{9} and this material is superconducting
\cite{6}. In comparison, the DOS at Fermi level for K$_8$Ge$_{46}$ is moderate
and it is not superconducting.

\section{Conclusions}

We have used first principles SCF pseudopotential method to investigate
the structural and electronic properties of Ge$_{46}$ and K$_8$Ge$_{46}$
clathrates. The main conclusion of this work can be made as follows:

(1) Germanium clathrate Ge$_{46}$ is found to be a locally stable structure
with its energy only slight higher than that of diamond phase, and its atomic
volume is about 13$\%$ larger than diamond phase.

(2) Ge$_{46}$ clathrate shows an indirect band gap along $\Gamma$-$X$
direction that is about 1 eV higher than the band gap of diamond phase.
The pentagonal rings in the clathrate structure cause the valence band
structure of Ge$_{46}$ clathrate to be similar to that of Ge$_{24}$
fullerene cage. The open covalent network structures contribute to
the large band gaps.

(3) The K$_8$Ge$_{46}$ clathrate is metallic with a moderate density of
states. The valence band structures and DOS are similar to those of the
pure Ge$_{46}$, while the conduction bands are modified due to the K
dopants. Almost complete charge transfer from K sites to Ge frames is found
in the K$_8$Ge$_{46}$ clathrate.

\begin{acknowledgements}
This work is supported by the U.S. Army Research Office
(Grant DAAG55-98-1-0298) and Department of Energy (Grand DEFG02-96ER45560).
The authors thank O.Zhou for helpful
discussions. We acknowledge computational support from the
North Carolina Supercomputer Center.
\end{acknowledgements}

\end{document}